\documentclass[usegraphicx,useAMS,usenatbib]{mn2e}

\pubyear{2007}

\begin{document}

\title[Probing Dark Matter with Pulsars]
{Probing Dark Matter Substructure with Pulsar Timing}

\author[E. R. Siegel et al.]
{E. R. Siegel,$^{1,2}$\thanks{E-mail: esiegel2@wisc.edu} M. P.
Hertzberg$^{3}$ and J. N. Fry$^{2}$ \\
$^1$Department of Physics, University of Wisconsin, Madison, WI,
53706, USA \\
$^2$Department of Physics, University of Florida,
Gainesville, FL, 32611-8440, USA \\
$^3$Center for Theoretical Physics, Massachusetts Institute of
Technology, Cambridge, MA, 02139, USA}

\date{Accepted ???? ?????? ??.  Received ???? ?????? ??; in original form 2007 February 19}

\maketitle

\begin{abstract}
We demonstrate that pulsar timing measurements may potentially be
able to detect the presence of dark matter substructure within our
own galaxy.  As dark matter substructure transits near the
line-of-sight between a pulsar and an observer, the change in the
gravitational field will result in a delay of the
light-travel-time of photons.  We calculate the effect of this
delay due to transiting dark matter substructure and find that the
effect on pulsar timing ought to be observable over decadal
timescales for a wide range of substructure masses and density
profiles. We find that transiting dark matter substructure with
masses above $10^{-2} \, M_\odot$ ought to be detectable at
present by these means.  With improved measurements, this method
may be able to distinguish between baryonic, thermal non-baryonic,
and non-thermal non-baryonic types of dark matter.  Additionally,
information about structure formation on small scales and the
density profiles of galactic dark matter substructure can be
extracted via this method.
\end{abstract}


\section{Introduction \label{I}}

One of the most fundamental goals of cosmology is to determine the
composition of the universe.  Contrary to our experience in the
solar system, where the emitted light traces the mass
distribution, the vast majority of matter in the universe appears
to be non-luminous.  This puzzle is known as the dark matter
problem.  One of the greatest goals in modern cosmology is to
determine the nature and properties of this dark matter. Although
many solutions have been proposed explaining the presence of dark
matter, its nature remains mysterious, as all but the most
indirect methods of detection have been unsuccessful thus far.

The astrophysical evidence for the existence of dark matter is
overwhelming, as recently reviewed in \citet{Siegel:06}.  While
the amount of matter in luminous sources is only $\Omega_\star
\simeq 0.005$ \citep{FP:04,Salucci:06}, precision measurements of
the cosmic microwave background \citep{Spergel:06} and type Ia
supernovae \citep{Riess:06} indicate that the total amount of
matter is $\Omega_m \simeq 0.26$, or that $98$ per cent of the
matter in the universe is non-luminous.  While a significant
component of this non-luminous matter can be baryonic,
nucleosynthesis \citep{bbn}, Silk damping effects \citep{SilkLSS},
and searches for MAssive Compact Halo Objects (MACHOs)
\citep{EROS} constrain the amount of baryonic matter to be
$\Omega_b \approx 0.04$. This means that, although a portion of
the dark matter is baryonic $(\sim 13 \, \mathrm{per cent})$, the
vast majority of matter in the universe is both non-luminous and
non-baryonic $(\Omega_\mathrm{DM})$.
This non-baryonic dark matter must also be cold; i.e., it must
have become non-relativistic at early times, as evidenced most
clearly by the large amount of power on small scales at early
times in the Lyman-$\alpha$ forest \citep{Lya}. This constraint
eliminates the only standard-model candidate for non-baryonic dark
matter, the neutrino.  There are then two generic classes of
candidates for cold, non-baryonic dark matter: the thermal relic
and the non-thermal relic.  A thermal relic is a particle produced
by thermally efficient interactions in the background plasma of
the early universe.  Its abundance then freezes out as the
temperature drops, but it remains in kinetic equilibrium with the
background plasma until later times \citep{Edbert:06}.  Examples
of thermal relic dark matter candidates include the lightest
supersymmetric particle, lightest Kaluza-Klein particle, and other
particles generically classified as Weakly Interacting Massive
Particles (WIMPs).  On the other hand, non-thermal relics can be
produced at phase transitions by vacuum misalignment mechanisms,
and are born non-relativistic. Examples of non-thermal dark matter
candidates include axions \citep{PQ,Weinberg,Wilczek} and massive
gravitons \citep{Tinyakov}.  Finally, it is possible that both
thermal and non-thermal types of dark matter exist, and that
$\Omega_\mathrm{DM}$ is made up of a composite of these two
classes.


In a recent and exciting development, non-baryonic dark matter has
been detected indirectly via its gravitational influence in the
Bullet Cluster \citep{bullet}. Since baryons, thermal relic dark
matter, and non-thermal dark matter all decouple from the
primordial plasma at different times and possess different
interactions, it is expected that the small-scale structure that
forms composed of these types of matter will be quite different
from one another. Baryons will remain in either diffuse gas clouds
or collapse to form MACHOs, thermal dark matter will have its
small scale structure on scales below $10^{-4}$ to $10^{-6} \,
M_\odot$ suppressed by Silk damping, whereas non-thermally
produced dark matter will have no such cutoff in its spectrum.
Thermal dark matter will produce, on small scales, diffuse WIMP
microhalos after gravitational collapse down to masses of about
one Earth mass \citep{Green,LZ,Diemand,Edbert:06}, whereas
non-thermal dark matter will produce collapsed structure down to
much smaller mass scales, without a cutoff in the spectrum, which
will produce Non-thermal Axionic Collapsed HalOS (NACHOs).  Under
certain circumstances \citep{Sikivie:06,Zurek}, the non-thermal
dark matter will produce halos with substantial enhancements in
density and abundance at $ \sim 10^{-12} \, M_\odot$, known as
axion miniclusters. In either case, these dark matter structures
on the smallest scales may survive intact to the present day as
substructure within larger collapsed structure such as our own
galaxy \citep{Gnedin:06}. While large-scale dark matter structures
like our galaxy's halo are expected to be smooth to a simple
approximation, realistic models include clumps, cusps, and
possibly caustics.

This paper demonstrates that pulsar timing measurements can be
used to probe the dark matter substructure in our own galaxy which
transits across the line-of-sight between a pulsar and an
observer.  As the dark matter substructure transits, the variation
in the light-travel-time of the pulses is of sufficient magnitude
that the more stable millisecond pulsars should be able to not
only detect the substructure, but to discriminate between the
different signatures of MACHOs, NACHOs, and WIMPs. The remainder
of this paper is layed out as follows.  Section \ref{II} presents
a derivation of the delay in the light-travel-time due to a
gravitational source intervening near or across the line-of-sight
between an emitter and an observer, providing a number of
examples.  Section \ref{III} applies this idea to the physical
case of using pulsar timing measurements to detect the intervening
dark matter substructure between Earth and the pulsar of interest.
Section \ref{IV} discusses the possibility of differentiating
between different types of dark matter and constraining its
properties through precision timing measurements.  Finally,
Section \ref{V} summarizes the major successes and drawbacks of
this method, detailing the challenges of using pulsar timing to
search for dark matter in our own galaxy.

\section{Gravitational Time Delay \label{II}}

One consequence of general relativity is the delay in the
light-travel-time (l.t.t.) due to intervening gravitational
sources between an emitter and an observer \citep{Shapiro:64}.
This effect, known (for a point source) as the Shapiro time delay,
is one of the classic solar system tests of Einstein's relativity
\citep{grqc0103044}. This effect can be applied to extended
sources and larger scales as well.  If any collapsed substructure
within our galaxy transits across the line-of-sight (l.o.s.)
between an emitter and an observer, then the l.t.t. will be
altered by the gravitational presence of the galactic
substructure.  The remainder of this Section is devoted to a
derivation of the magnitude of the delay in the l.t.t. due to
transiting collapsed substructure.

Consider the configuration shown in Figure 1: an emitting source
is a fixed distance $l$ from an observer, with a transiting
gravitational source near the l.o.s., with a minimum distance to
the l.o.s. (i.e. impact parameter) $b$. The gravitational source
has a velocity $\vec{v}(t)$ and position $\vec{x}(t)$, where the
velocity, assumed to be constant, has components parallel
$(v_{\parallel})$ and perpendicular $(v_\perp)$ to the l.o.s.,
while the position deviates from its initial value only in the
direction given by $\vec{v}$. The two cases of physical importance
to consider are when the radial size of the gravitational source
$(r_\mathrm{max})$ is either greater or smaller than the impact
parameter, $b$.
\begin{figure}
\label{figorig}
\includegraphics[width=\columnwidth]{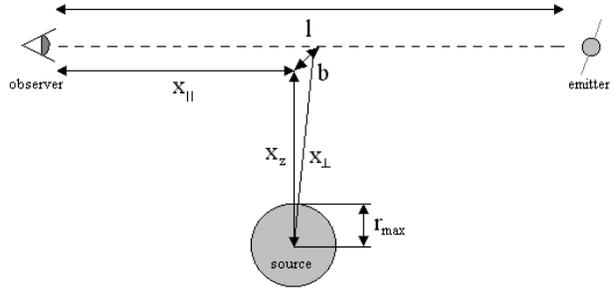}
\caption{The generic configuration for a fixed emitter and
observer, with a gravitational source of radius $r_\mathrm{max}$
transiting across or close to the line-of-sight.  The
emitter-observer distance is taken to be $l$, while the
gravitational source has a distance to the observer $\vec{x}$ and
a velocity $\vec{v}$, both of which have components perpendicular
$({x}_\perp$, ${v}_\perp)$ and parallel $({x}_\parallel$,
${v}_\parallel)$ to the line-of-sight. The minimum distance
between the gravitational source and the line-of-sight as the
source transits is defined as the impact parameter, $b$. }
\end{figure}

To determine the time delay induced by the presence of this
gravitational source, we integrate a null geodesic from the
emitter to the observer.  We consider the weak-field metric for a
single gravitational source,
\begin{equation}
\label{metric} ds^2 = - (1 + 2 \frac{\phi}{c^2}) c^2 dt^2 + (1 - 2
\frac{\phi}{c^2}) d\chi^2 + \chi^2 d \Omega ^2 \mathrm{,}
\end{equation}
where $\phi$ is the gravitational potential induced by the
presence of a single gravitational source. In the presence of
multiple sources, we note that this procedure can be used to
calculate the time delay for each source individually and then the
effects can be summed, since the fields are weak.

We choose a radial null geodesic along the l.o.s. from the
emitting source to the observer, working in coordinate time, and
find that the total l.t.t. is
\begin{equation}
\label{ltt} t = t_0 + \Delta t =
\int_\mathrm{emit}^\mathrm{observe} dt =
\int_\mathrm{emit}^\mathrm{observe} \frac{1}{c} (1 - 2
\frac{\phi}{c^2}) d\chi \mathrm{.}
\end{equation}
The first term on the right-hand-side of equation (\ref{ltt})
integrates to give a constant $(l / c)$ which we identify as
$t_0$, or the l.t.t. in the absence of any gravitational sources.
The second term is indicative of the time delay $(\Delta t)$ due
to the presence of a gravitational source.

Since the gravitational fields of interest are weak $(\phi / c^2
\ll 1)$, it is an excellent approximation to treat $\phi$ as a
Newtonian potential when outside the gravitational source, $\phi =
- G M / r$.  When one is inside a spherical gravitational source,
one must be a little more careful and perform the following
integral:
\begin{equation}
\label{phiinside} \phi = - \int_0^r d^3r' \frac{G \, \rho(r')}{r}
- \int_r^{r_\mathrm{max}} d^3r' \frac{G \, \rho(r')}{r'}
\mathrm{,}
\end{equation}
which reduces to $\phi = - G M / r$ at $r=r_\mathrm{max}$. In
general, the effect of the time delay can then be quantified by
the integral from the emitter to the observer
\begin{equation}
\label{Shapiro} \Delta t = \int_0^l \frac{ - 2 \phi(x') }{c^3} dx'
\mathrm{,}
\end{equation}
where the $r$ from equation (\ref{phiinside}) is the distance
between each given position along the l.o.s. $(x')$ and the
center-of-mass of the gravitational source. 
For points lying outside of the source, the total mass
$(M_\mathrm{tot})$ of the source contributes to $\phi$, whereas
the gravitational potential can either increase or decrease as one
penetrates the source, dependent upon the density profile.

For an extended source of radius $r_\mathrm{max}$, the enclosed
mass within the source is given by
\begin{equation}
\label{Menc} M_\mathrm{enc}(x') =
\int_\mathrm{0}^{\sqrt{(x'-x_\parallel)^2 + (x_\perp)^2}} 4 \pi
r^2 \rho(r) dr \mathrm{,}
\end{equation}
but merely substituting $M_\mathrm{enc}$ for $M_\mathrm{tot}$ is
insufficient; one must recalculate $\phi$ at each point and then
solve equation (\ref{Shapiro}).  
The total mass of the gravitational source $(M_\mathrm{tot})$ is
given by equation (\ref{Menc}) at the point where $r_\mathrm{max}
= \sqrt{(x'-x_\parallel)^2 + (x_\perp)^2}$. In general, therefore,
the Shapiro time delay will be given by the combination of
equation (\ref{Shapiro}) with the appropriate value for $\phi$ at
all points.

We first consider the case of impact parameters, $b$, that are
larger than the maximal radius $(r_\mathrm{max})$ of the
gravitational source in question.  This approximation corresponds
to the physical cases of baryonic substructures (stars and MACHOs)
and for non-baryonic substructure where the object is sufficiently
far away, even at its point of closest approach, so that its
radius never intersects the l.o.s. between the emitter and
observer.  For this case $( b > r_\mathrm{max} )$, the enclosed
mass is always $M_\mathrm{tot}$, and equation (\ref{Shapiro}) is
exactly integrable, yielding a time delay of
\begin{equation}
\label{out} \Delta t = \frac{2 G M_\mathrm{tot}}{c^3} \ln {\left(
\frac{l - x_\parallel + \sqrt{ (l-x_\parallel)^2 + (x_\perp)^2 }}{
- x_\parallel + \sqrt{ (x_\parallel)^2 + (x_\perp)^2 }} \right)}
\mathrm{.}
\end{equation}
In the region of physical interest, where the approximation $\{
l,x_\parallel, l - x_\parallel \} \gg \{ b,x_\perp \}$ is valid, a
simpler expression of
\begin{equation}
\label{outside} \Delta t \simeq \frac{2 G M_\mathrm{tot}}{c^3} \ln
{\left( \frac{4 x_\parallel (l - x_\parallel)}{ (x_\perp)^2 }
\right)}
\end{equation}
can be obtained.  Numerically, this yields a time delay dependent
only on the parameters $l$, $\vec{x}$, and $M_\mathrm{tot}$, such
that
\begin{equation}
\label{Noutside} \Delta t \simeq 2.95 \times 10^{-11} \,
\mathrm{s} \left( \frac{M_\mathrm{tot}}{M_\oplus} \right) \, \ln
{\left( \frac{4 x_\parallel (l - x_\parallel)}{ (x_\perp)^2 }
\right)} \mathrm{,}
\end{equation}
where $M_\oplus$ is one Earth mass.  As the position of the
gravitational source, $\vec{x}$, changes with time, so too will
the time delay, $\Delta t$.  We note that the time delay of
equations (\ref{out} - \ref{Noutside}) depends very weakly on
changes in $l$ and $x_\parallel$, but is more sensitive to changes
in $x_\perp$, since $x_\perp \ll \{ x_\parallel , l - x_\parallel
\}$.  We also note that these results indicate that the time delay
scales linearly with mass, but only logarithmically with position.

\begin{figure}
\label{fig1}
\includegraphics[width=\columnwidth]{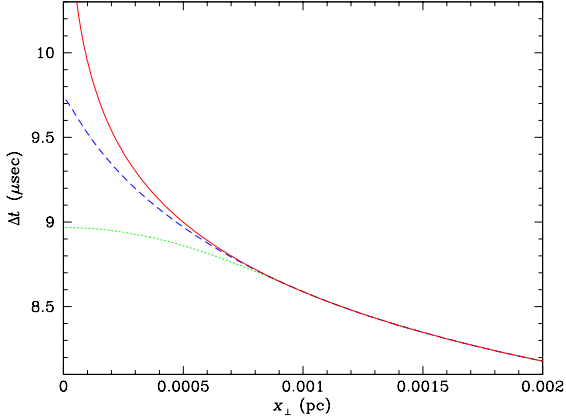}
\caption{$\Delta t$ vs. $x_\perp$ with $l = 2 \, \mathrm{kpc}$,
$x_\parallel = 1 \, \mathrm{kpc}$ and $M_\mathrm{tot} = 10^4 \,
M_\oplus$ for three sources: a point source (solid line), a
constant density halo (dotted line), and an NFW profile (dashed
line). Additionally, the physical size of the gravitational source
for extended objects, $r_\mathrm{max}$, is chosen to be $10^{-3}
\, \mathrm{pc}$.  The NFW profile illustrated has a turnover
radius $r_0 \simeq 10^{-4.5} \, \mathrm{pc}$. $x_\perp$ is plotted
in units of parsecs, $\Delta t$ in units of seconds. This result
scales linearly with mass, as indicated in equation
(\ref{Noutside}), so that a microhalo $100$ times as massive would
have a change in
time delay $100$ times as large. 
}
\end{figure}

As the gravitational sources are expected to be extended objects
if they are non-baryonic, we also consider the case that the
l.o.s. intersects the gravitational source, so that $b <
r_\mathrm{max}$. The general equation will still be given by
equation (\ref{Shapiro}), although the integration will be
significantly more difficult here. Generically, the integral must
be broken up into three parts: from the observer to the near edge
of the source, from the near edge through the source to the far
edge, and from the edge of the source to the emitter. The integral
of equation (\ref{Shapiro}) then becomes
\begin{eqnarray}
\label{inside1} \Delta t &=& \int_0^{x_\parallel -
\sqrt{(r_\mathrm{max})^2 - (x_\perp)^2}} \left( \frac{ 2 G
M_\mathrm{tot} }{c^3 r} \right) dx' \nonumber\\
&+& \int_{x_\parallel - \sqrt{(r_\mathrm{max})^2 -
(x_\perp)^2}}^{x_\parallel + \sqrt{(r_\mathrm{max})^2 -
(x_\perp)^2}} \left( \frac{ - 2 \phi(x') }{c^3} \right) dx' \nonumber\\
&+& \int_{x_\parallel + \sqrt{(r_\mathrm{max})^2 - (x_\perp)^2}}^l
\left( \frac{  2 G M_\mathrm{tot} }{c^3 r} \right) dx' \mathrm{,}
\end{eqnarray}
where $\phi(x')$ is given by equation (\ref{phiinside}), and is
valid for any generic density profile $\rho(r)$ for the
gravitational source.  An oftentimes more useful form for equation
(\ref{inside1}) is the partially solved equation
\begin{eqnarray}
\label{inside2} \Delta t &\simeq& \frac{2 G M_\mathrm{tot}}{c^3}
\ln {\left[ \frac{4 x_\parallel (l - x_\parallel)(r_\mathrm{max} -
\sqrt{(r_\mathrm{max})^2 - (x_\perp)^2})}{ (x_\perp)^2
(r_\mathrm{max} + \sqrt{(r_\mathrm{max})^2 - (x_\perp)^2})}
\right] }
\nonumber\\
&+& \int_{x_\parallel - \sqrt{(r_\mathrm{max})^2 -
(x_\perp)^2}}^{x_\parallel + \sqrt{(r_\mathrm{max})^2 -
(x_\perp)^2}} \frac{- 2 \phi(r)}{c^3 } d^3 r
\mathrm{,}
\end{eqnarray}
which contains the approximations $x_\perp \ll \{x_\parallel,
l-x_\parallel\}$, but still admits a generic density profile
$\rho(r)$ (implicit in $\phi(r)$). For all physical cases (i.e.,
$\rho(r)$ is everywhere non-negative), this leads to a smaller
time delay compared to the case of a point mass for sufficiently
small values of $x_\perp$. Figure \ref{fig1} illustrates $\Delta
t$ as a function of $x_\perp$ for the case of a point source, the
case of a constant density halo with $\rho(r) = 3 M_\mathrm{tot} /
4 \pi r_\mathrm{max}^3$, as well as the case of a halo with an NFW
\citep{NFW:95,NFW:97} profile with a cutoff at $r_\mathrm{max}$,
obeying the formula
\begin{equation} \label{NFWprofile}
\rho(r) = \frac{M_\mathrm{tot}}{4 \pi r (r_0 + r)^2 \left[ \ln{(1
+ \frac{r_\mathrm{max}}{r_0})} - \frac{r_\mathrm{max}}{r_0 +
r_\mathrm{max}} \right]} \mathrm{,}
\end{equation}
where $r_0$ is the turnover radius between an outer halo density
falling off as $1 / r^3$ and and inner halo density which falls as
$1 / r$.  As can be seen in Figure \ref{fig1}, smaller impact
parameters are more sensitive to the form of a gravitational
source's density profile.

\section{Dark Matter Detection \label{III}}


The analysis in Section \ref{II} has demonstrated that a
gravitational source lying near the l.o.s. between an emission
source and an observer will cause a delay in the l.t.t. of the
emitted signal.  We now specifically turn to the case of our own
galaxy for a physical application.  It is not possible to measure
the absolute l.t.t. of any astrophysical object directly, but {\it
changes} in the l.t.t. can be observed over time.  To measure a
change in the l.t.t., we require a stable, predictable emitter
whose emission properties remain consistent over long timescales.
The most stable, consistent emitters in the known universe are the
millisecond pulsars, many of which have remained stable without
glitches, starquakes, or flux unpinning over timescales exceeding
thirty years.  Pulsars have been used in many instances in the
past to glean information about the physical universe
\citep{Det:79}. By treating millisecond pulsars as emission
sources, the l.t.t. of each pulse as it heads towards Earth will
be quite sensitive to any changes in the matter distribution near
the l.o.s.  Thus, sufficiently accurate measurements of the l.t.t.
will allow us to search for all massive transiting objects near
the l.o.s., including MACHOs, WIMPs, and NACHOs.

Pulsars are the ideal candidates to serve as emission sources, as
they are the most perfect natural clocks in the known universe.
Pulsars exist with a wide variety of rotational periods and also
in the emission periods of their pulses.  The pulsars with the
highest rotational frequencies, and hence the shortest
pulse-to-pulse periods, also happen to be the most stable types of
pulsars, with a period $(T_p)$ of $\mathcal{O}(1 \, \mathrm{ms})$.
These millisecond pulsars have typical residuals of $\mathcal{O}(1
\, \mu \mathrm{s})$, meaning that if one measures enough pulses
accurately (typically over $\sim 1 \, \mathrm{hr}$ timescales),
the deviation of any pulse-to-pulse measurement from the average
value is $\mathcal{O}(1 \, \mu \mathrm{s})$. Remarkably, these
residual uncertainties are {\it not cumulative}, so that the
arrival-time uncertainty between the first pulse and the $n$-th
pulse remains $\mathcal{O}(1 \, \mu \mathrm{s})$, and does not
increase as the number of pulses increases.  Because of these
properties of millisecond pulsars, the ones which are completely
stable (no glitches, starquakes, or flux unpinning) have their
periods known to an accuracy of $\sim 10^{-17} \, \mathrm{s}$ or
better, as an $\mathcal{O}(1 \, \mu \mathrm{s})$ uncertainty over
a span of $\mathcal{O}(10^{12})$ pulses (the first millisecond
pulsar was discovered in 1982) yields a measured period accurate
to $\sim 10^{-18} \, \mathrm{s}$.

\begin{figure}
\label{Mark2}
\includegraphics[width=\columnwidth]{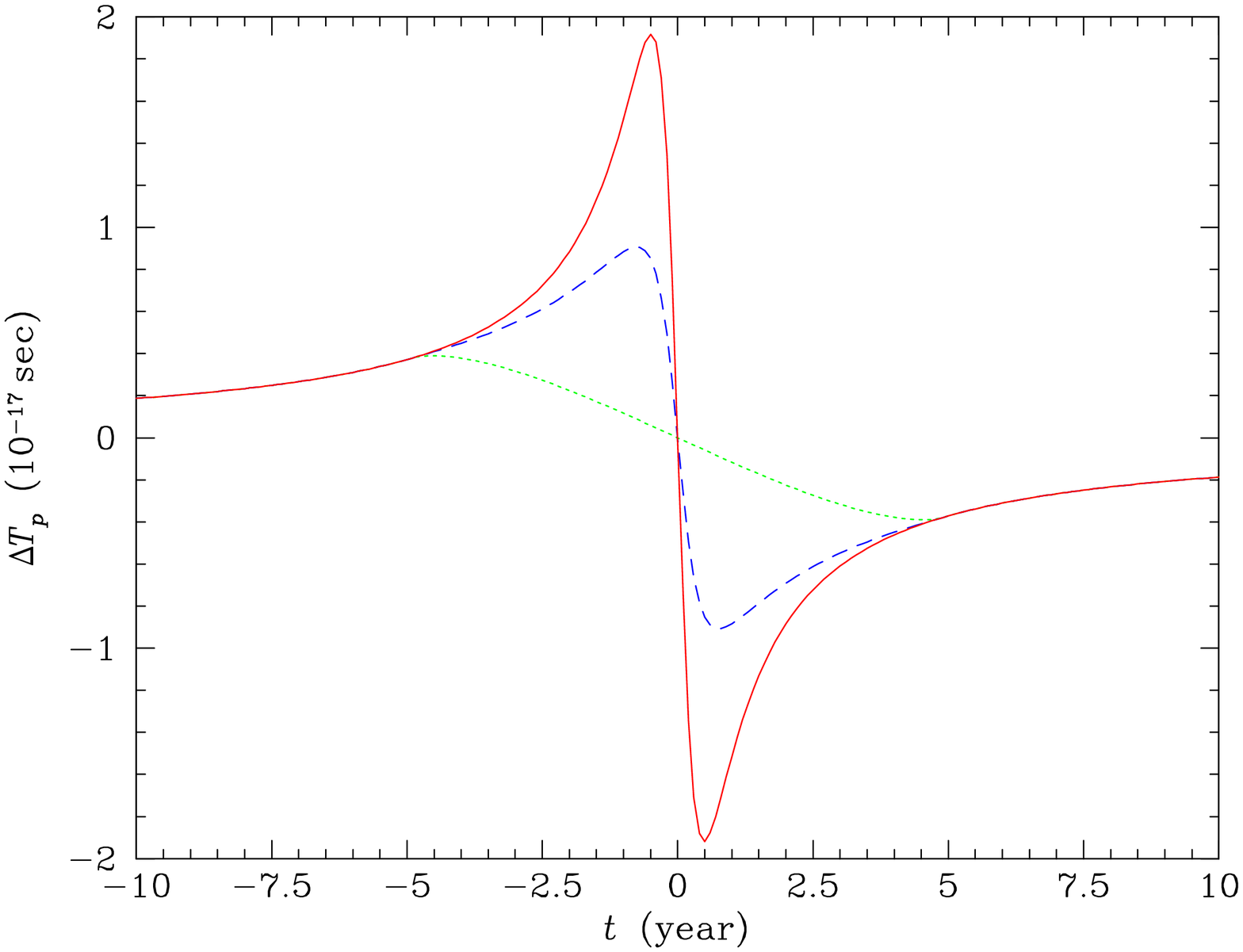}
\caption{$\Delta T_p$ vs. time for a set of parameters
corresponding to the variables defined in Figure \ref{figorig}: $l
= 2 \, \mathrm{kpc}$, $x_\parallel = 1 \, \mathrm{kpc}$,
$M_\mathrm{tot} = 10^4 \, M_\oplus$, $b=10^{-4} \, \mathrm{pc}$,
$v_\perp = 200 \, \mathrm{km} \, \, \mathrm{s}^{-1}$, and
$T_p=10^{-3}\mathrm{s}$.  $\Delta T_p$ is illustrated in seconds
and time in years.  The solid curve is for a point mass, the
dotted curve is for a microhalo of constant density with size
$r_\mathrm{max}= 10^{-3} \, \mathrm{pc}$, and the dashed curve is
for an extended microhalo with an NFW profile of size
$r_\mathrm{max}= 10^{-3} \, \mathrm{pc}$ and turnover radius $r_0
= 10^{-4.5} \, \mathrm{pc}$. 
}
\end{figure}

In order to detect the time delay induced by transiting
substructure, the effect must, in some sense, be large when
compared with the intrinsic timing uncertainties.  The uncertainty
in the pulse arrival time is primarily due to environmental
uncertainties, owing to the difficulty in modelling both our own
and the pulsar's local environments.  In principle, the simplest
method for detecting substructure gravitationally would be to
determine the shift in the expected pulse arrival time $(\Delta
T_p)$ for adjacent pulses due to changes in the gravitational
potential near the l.o.s.  This can be computed by examining the
shift in $\Delta t$ over the time of one pulse, $T_p$, as the
intervening matter changes position.  Assuming a constant velocity
for the intervening gravitational source, $\Delta T_p$ is
calculable in general from equation (\ref{inside2}) by taking the
difference in $\Delta t$ between an initial source position
$\vec{x}_i = (b, x_\parallel, x_\perp)$ and a final source
position $\vec{x}_f = (b, x_\parallel + v_\parallel T_p, x_z +
v_\perp T_p)$.  So long as $\Delta T_p$ is greater than the known
period accuracy $(\sim 10^{-17} \, \mathrm{s})$, it may be
possible to extract the signal that would be caused by transiting
substructure using statistical techniques, despite the large
pulse-to-pulse residuals. $\Delta T_p$ as a function of time is
shown in Figure \ref{Mark2} for the three transiting cases of a
point mass, a constant density extended object, and an extended
microhalo with an NFW profile. Unfortunately, this method is not
presently feasible in practice for the detection of masses
significantly below $1 \, M_\odot$, since pulse arrival times have
not been measured to accuracies even approaching $10^{-17} \,
\mathrm{s}$.

A more complicated, but perhaps more feasible method would be to
integrate the observed and expected time delays over many pulses.
For this method, we are not constrained by the timescale $T_p$, as
we are free to measure $\Delta T_p$ over any time interval that is
a multiple of $T_p$.  This allows us to enhance the magnitude of
the signal from the pulse-to-pulse values of Figure \ref{Mark2} up
to the difference in the time delay between two given positions.
The method of computing the change in the l.t.t. is the same as
above, except $T_p$ in the final position is to be replaced by $n
T_p$, where $n$ is any positive integer.  While the noise residual
does not decrease for this case as more pulses are observed, it
does not increase either, as the $\mathcal{O}(1 \, \mu
\mathrm{s})$ uncertainty is a non-cumulative error. Therefore, as
long as the change in the time delay over a reasonably measurable
timescale is comparable to or greater than the residual
uncertainties, this signal will be detectable {\it in practice}.
Calculations for the integrated effect of the time delay over many
years for a transiting point mass, extended constant density halo,
and an extended halo with an
NFW profile are illustrated in Figure 4.
\begin{figure}
\label{awesome}
\includegraphics[width=\columnwidth]{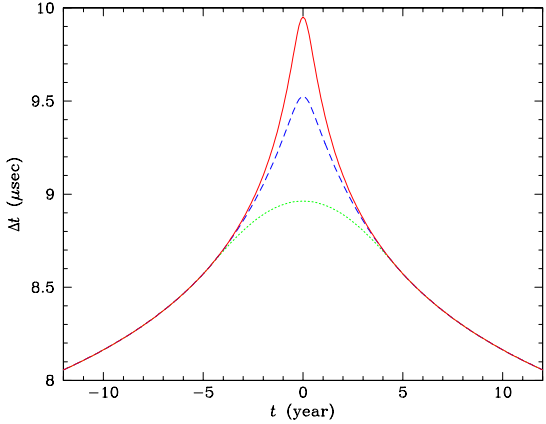}
\caption{ $\Delta t$ vs. time for a set of parameters
corresponding to the variables defined in Figure \ref{figorig}: $l
= 2 \, \mathrm{kpc}$, $x_\parallel = 1 \, \mathrm{kpc}$,
$M_\mathrm{tot} = 10^4 \, M_\oplus$, $b=10^{-4} \, \mathrm{pc}$,
$v_\perp = 200 \, \mathrm{km} \, \, \mathrm{s}^{-1}$, and
$T_p=10^{-3}\mathrm{s}$.  $\Delta T_p$ is illustrated in seconds
and time in years.  The solid curve is for a point mass, the
dotted curve is for a microhalo of constant density with size
$r_\mathrm{max}= 10^{-3} \, \mathrm{pc}$, and the dashed curve is
for an extended microhalo with an NFW profile of size
$r_\mathrm{max}= 10^{-3} \, \mathrm{pc}$ and turnover radius $r_0
= 10^{-4.5} \, \mathrm{pc}$.  The chosen set of parameters is very
optimistic, but demonstrates the potential power of this method.
}
\end{figure} Quantitatively, for a residual of $\mathcal{O}(1 \,
\mu \mathrm{s})$, a transiting mass of $M_\mathrm{tot} \approx
10^4 \, M_\oplus$ with an impact parameter of $b \approx 10^{-4}
\, \mathrm{pc}$ will be at the limit of detectability.  Detection
is made more difficult by the fact that modern pulsar analysis
techniques require the removal of linear and quadratic terms from
millisecond pulsar data before any further analysis is done. The
majority of close dark matter encounters (excepting actual
observed transits) will have their signatures absorbed in this
subtraction. We suggest, therefore, that it may be beneficial to
develop non-standard analysis techniques to explicitly search for
this effect.

As can be seen from Figures \ref{Mark2} and 4, the time delay
signal is very sensitive to the density profile of the transiting
source, and can be used to determine the densities of sufficiently
massive transiting dark matter, if their impact parameters are
small enough. Halos with steeper density profiles and smaller
cores are more easily detectable, whereas the more diffuse ones
will have a more significant departure from the point-source
template.  Of course, at sufficently large impact parameters, all
source behave as point sources.  Many of the data analysis
techniques being pioneered by groups searching for a cosmic
gravitational wave background with pulsars \citep{GWgroup} should
be applicable to identifying the time delay signature of
transiting dark matter substructure.  As a caveat, it is worth
pointing out that unless the cusp of the Shapiro delay is clearly
visible in the residuals, the fit is likely to become degenerate
in the presence of noise, as fast-transiting small masses with
large impact parameters will be difficult to discern from
slow-moving massive objects with small impact parameters.

\section{Dark Matter Discrimination \label{IV}}

The small-scale structure produced in our universe is very
sensitive to the type (thermal, non-thermal, or baryonic) and
particle properties (masses and couplings) of dark matter.  While
baryons cannot be all of the dark matter, they do compose a
significant amount of the nonluminous matter in the universe
($\sim 13 \, $per cent).  Baryons do not exist in isolation in
small mass clumps, but rather as a part of very large mass
structures, due to the fact that their collapse is suppressed on
all scales which enter the horizon prior to baryon-photon
decoupling (at recombination). Therefore, even though baryons
exist in the form of diffuse gas clouds and compact halo objects
today (in addition to stars), these baryons were never isolated
from the non-baryonic dark matter halos in which they are found.
The signature that baryonic dark matter (MACHOs) will leave in the
pulsar timing measurements is that of a point mass source, as
given by changes over time in equation (\ref{Noutside}).  For
negligible changes in $l$ and $x_\parallel$, this would result in
a pulse-arrival-time shift of
\begin{equation}
\label{Machodelay} \Delta t_f - \Delta t_i \simeq 2.95 \times
10^{-11} \, \mathrm{s} \, \left( \frac{M_\mathrm{tot}}{M_\oplus}
\right) \ln{\left( \frac{b^2 + x_z^2}{b^2 + (x_z + v_\perp n
T_p)^2} \right)}
\end{equation}
between $n$ pulses. Additionally, MACHOs are compact enough that
effects such as gravitational microlensing ought to be observable
if one knows where to look as well, allowing for possible
cross-correlation effects.

The vast majority ($\sim 87 \, $per cent) of dark matter is
non-baryonic, however (contained in $\Omega_\mathrm{DM}$).  In
contrast to baryons, the non-baryonic matter is practically
collisionless, and thus collapses to form much more diffuse
structures than ordinary matter.  Additionally, the suppression of
non-baryonic structure ceases at a much earlier epoch, as thermal
dark matter kinetically decouples from the primordial plasma at a
typical temperature of $\mathcal{O}(10 \, \mathrm{MeV})$, and
non-thermal dark matter is always decoupled both thermally and
kinetically from the plasma. The exact epoch of kinetic decoupling
of thermally produced dark matter is determined by the mass of the
decoupled particles and its interactions; this in turn determines
the mass function of small-scale dark matter clumps
\citep{Edbert:06}.  As a result, thermally produced dark matter
will have a sharp drop in its mass function for masses below $\sim
20 \, M_\oplus$, with the exact value dependent upon the exact
particle properties chosen. Non-thermal dark matter, by contrast,
will not only have no such feature, but may also have a sharp
enhancement in the number density of collapsed structures on very
small scales under the right conditions.  For the case of
non-thermally produced axions where the Peccei-Quinn symmetry is
broken subsequent to the end of inflation, the initial density
fluctuations on scales corresponding to the horizon at $T \simeq
\Lambda_\mathrm{QCD}$ will be greatly enhanced \citep{Sikivie:06}.
This will result in much more numerous and denser structures on
these small scales ($M \sim 10^{-6} M_\oplus$) as compared with
thermal dark matter, although they are still quite diffuse as
compared to baryons. While mass scales of $10^{-6} M_\oplus$ are
far beyond the reach of any foreseeable technology using pulsar
timing, this density enhancement will affect larger mass scales as
well, and may be detectable if there is a substantial enhancement
at mass scales which can be probed, either today or in the future,
by pulsar timing.

The surefire way to distinguish between the various types of dark
matter is to probe the low end of the mass function, where WIMPs
and NACHOs differ from one another in their abundances.  Thermal
relic dark matter begins to experience a sharp dropoff in the
number of collapsed structures at masses below $\sim 20 \,
M_\oplus$, although that number is model dependent.  If the
abundances of these dark matter microhalos can be measured with
sufficient accuracy at small mass scales, it may be possible to
reconstruct the low end of the mass function, thus constraining
the types and properties of dark matter.  It is also of interest
to probe the density profiles of these structures, which may shed
light on not only whether axion miniclusters are present in the
universe, but also on the nonlinear collapse of small-scale
structure.

In order to calculate the types of observations necessary to probe
the effect of transiting dark matter, we must calculate the
expected transit rate of these dark microhalos.  Moreover, as it
is the mass function we desire to probe, different halo masses
must be considered and searched for.  Press-Schechter theory
\citep{PS:74} indicates that the number density of collapsed
structures which should be found in our universe today is strongly
dependent upon the mass of the structure of interest.  At small
scales, the mass function is more accurate if replaced by the
Sheth-Tormen function \citep{ST:99}. The mass function $dn / dM$
then obeys the equation
\begin{equation}
\label{PS}
\frac{M}{\rho_m} \frac{dn}{d \, \ln{M}} = \frac{d \, (\ln{\nu})}{d
\, ( \ln{M} )} \nu f(\nu) \mathrm{,}
\end{equation}
with
\begin{equation}
\label{ST} \nu f(\nu) = 0.322 \left[ 1 + \left(
\frac{\nu^2}{\sqrt{2}} \right)^{-0.3} \right] \sqrt{\frac{\sqrt{2}
\nu^2}{\pi}} e^{- \frac{\nu^2}{2\sqrt{2}} } \mathrm{,}
\end{equation}
where $\nu = \sigma / \delta_c$, $\delta_c \simeq 1.686$, and
$\sigma$ is the root-mean-squared mass density perturbation in a
given region of space.  The mass density perturbation spectrum is
highly dependent upon the transfer function and power spectrum of
the dark matter, and to a lesser extent, the window function. It
is estimated in \citet{LZ} that there should be a significant
fraction of the Milky Way halo in dark matter clumps of
approximately $10^2 \, M_\oplus$, implying that there are
$10^{14}$ to $10^{15}$ of these clumps in our galaxy. (A more
accurate estimate of the mass function, although it assumes a
specific model of thermal dark matter, can be found in
\citet{Edbert:06}.) Assuming a pulsar-to-observer distance of $2
\, \mathrm{kpc}$, a typical transverse velocity of a dark matter
microhalo of $v_\perp \simeq 200 \, \mathrm{km} \, \,
\mathrm{s}^{-1}$, $10^{15}$ microhalos of mass $10^2 \, M_\oplus$,
and an NFW profile for the dark matter structure in our own galaxy
with a turnover radius of $25 \, \mathrm{kpc}$, the local number
density of dark matter microhalos is
\begin{equation}
\label{NmuH} n_{\mu \mathrm{H}} = \frac{\rho_{\mu
\mathrm{H}}}{10^{-4} \, M_\odot} \simeq 133 
\, \mathrm{pc}^{-3} \mathrm{,}
\end{equation}
based on a $\rho(r)$ for the Milky Way given by equation
(\ref{NFWprofile}) with $r_\mathrm{max} = 25 r_0$ and our radius
from the center of the galaxy taken to be $8 \, \mathrm{kpc}$. We
note that this estimate is at the high end of the possible number
density of microhalos, differing by as much as a factor of
$\mathcal{O}(10^2)$ from other estimates. This results in a
transit rate near the line-of-sight to a pulsar of
\begin{eqnarray}
\label{Trate} \Gamma \sim n_{\mu \mathrm{H}} \sigma v_\perp &=&
0.034 \left( \frac{n_{\mu \mathrm{H}}}{133 \, \mathrm{pc}^{-3}}
\right) \left( \frac{b_\mathrm{max}}{10^{-4} \, \mathrm{pc}}
\right) \left( \frac{l}{2 \, \mathrm{kpc}} \right) \nonumber\\
&& \times \left( \frac{v_\perp}{200 \, \mathrm{km} \,
\mathrm{s}^{-1}} \right) \frac{\mathrm{transits}}{\mathrm{year}}
\mathrm{,}
\end{eqnarray}
where $b_\mathrm{max}$ is the maximum impact parameter a microhalo
can have relative to the line-of-sight and still be of interest on
our observable $(\sim 25 \, \mathrm{years})$ timescales.  This
rate corresponds to approximately one microhalo with the above
properties transiting over the observed timescale of the pulsar.
The transit rates for different mass scales and number densities
can be computed in a similar fashion for differing mass functions.

The estimates given here are for dark matter microhalos embedded
in the galactic halo.  The presence of collapsed baryonic
substructure, however, particularly in the galactic disk, may
alter this profoundly.  Extended, low density astrophysical
objects such as dark matter microhalos may be gravitationally
disrupted (or even captured) by much more dense, massive baryonic
structure in the disk as the low density objects pass through
\citep{Gnedin:06}. On the other hand, this may not be the case, as
globular clusters, which are somewhat more extended and
significantly more massive, can remain intact and appear to be
undisturbed by the galactic disk \citep{Kobulnicky:05}. It is thus
unknown whether dark matter substructure will survive intact in
the galactic disk. There may therefore be a tremendous amount of
information obtainable by comparing the time-delay effects for
pulsars found in the galactic halo, where dark matter substructure
may be relatively undisturbed, with those found in the galactic
plane.

The net signal which ought to be searched for in the pulsar data
is a superposition of all the baryonic MACHOs and dark matter
microhalos which transit a significant distance in the $x_\perp$
direction with small impact parameters relative to the l.o.s. of
the pulsar of interest.  Based on the transit rate for low-mass
dark matter microhalos computed in equation (\ref{Trate}), it
ought to be a rare event to observe a transit with a large enough
mass $(\sim 10^{4} \, M_\oplus)$ to be detectable.  With
$\mathcal{O}(10)$ presently observed pulsars which are
sufficiently stable over their observed timescales of
$\mathcal{O}(10 \, \mathrm{years})$, there is a small but
reasonable probability that evidence for dark matter microhalos
exists in the current pulsar data.  Although there are over $100$
millisecond pulsars presently known, many of them are either weak
radio sources or located in globular clusters, rendering them
incapable of detecting dark matter through precision timing
campaigns with current instrumentation.  The transit rates are
small enough so that spatial averaging is not necessary, as only a
small number of these dark matter MACHOs (which behave as point
sources) and microhalos (which behave as extended sources) of
sufficient mass should be observable at all.

%
%
%

\section{Discussion \label{V}}

This paper has demonstrated the feasibility of using millisecond
pulsars to detect dark matter within our own galaxy.  Although the
dark matter is distributed in a smooth halo around our galaxy to a
zeroth approximation, departures from a smooth density
distribution (i.e., clumpiness, cuspyness, and caustics) are
features of realistic models of dark matter.  The dark matter
substructure found within our galaxy can be quantified through
theoretical modelling of small-scale structure through the deeply
nonlinear regime, but can also be probed experimentally through
precision measurements of pulsar timing. As the dark matter
distribution along the line-of-sight to a pulsar changes over
time, the light-travel-time changes as well (as given by equation
(\ref{Shapiro}) for an individual source).  This effect can be
quantified for each source, its effect on the pulse arrival time
can be calculated as detailed in Section \ref{III}, and summed
over all relevant transiting sources. We have illustrated that the
rate of interesting events is non-neligible, and such a signal may
already be present in archival data.  A successful detection of
dark matter substructure using pulsar timing would be the first
definitive detection of dark matter within our own galaxy.


Beyond merely detecting dark matter substructure, pulsar timing
has the potential to provide information about the masses and
density profiles of dark matter substructure.
Probing galactic substructure is the only method at present
capable of measuring the mass function on scales below $\sim 10^6
\, M_\odot$.  Accurate measurements of the mass function on small
scales can provide insights into dark matter particle properties
such as the type of dark matter (WIMPs or NACHOs), the temperature
of kinetic decoupling, and the dark matter's mass and scattering
cross sections.  Additionally, accurate measurements of the
density profile of these dark matter substructures, which is
directly related to the time delay signal to be probed by pulsar
timing, provides insights into both the particle properties and
the nonlinear evolution of dark matter. The density profiles may
be significantly different than standard cold dark matter
simulations indicate due to dark matter substructures interacting
with one another and with baryons, and due to the process of tidal
stripping, as analyzed in \citep{Gnedin:06}.  While the
survivability of these structures is still under debate, it
appears to be more of a quantitative question (i.e., what are the
percent of surviving halos, what percent are disturbed and to what
level, how much of the denser cores survive) than a qualitative
one.  If strong limits can be placed on the presence of dark
matter substructure in the galactic plane from pulsar timing, it
may be evidence indicating that tidal disruption of these
small-scale structures is very efficient.

In addition to the effects due to intervening dark matter, there
are two other interesting physical phenomena which can interfere
with this signal.  A gravitational wave background \cite{GWgroup}
is expected to have individual waves containing important effects
at the time delay level of $\mathcal{O}(1 \, \mathrm{ns})$, with
the overall stochastic background contributing a cumulative effect
of $\mathcal{O}(100 \, \mathrm{ns})$.  Although the signal has
specific templates which are quite different from that of
transiting dark matter, the pulse delay is of the same magnitude
for $\mathcal{O}(10^3 \, M_\oplus)$ substructure as it is for a
gravitational wave background, thus care must be taken to
distinguish these signals. Additionally, there are baryons in the
interstellar medium that can alter the pulse arrival time
\citep{Snesana:07}, although their effects may be correctable
\citep{You:07}. Although these signals are well-studied, they are
anticipated to have significantly larger effects on the pulse
arrival time, and thus they must be understood with great care in
order to be properly subtracted out.


The possibility of detecting dark matter using pulsar timing is
exciting and quite realistic.  With the transit rate given in
equation (\ref{Trate}) and the fact that there are $\sim 10$
pulsars with the necessary properties for detection, the limiting
factor is only to understand the pulsars' timing residuals. Recent
advances have allowed pulse-to-pulse residuals for some pulsars to
be constrained to $\mathcal{O}(0.1 \, \mu \mathrm{s})$ with $\sim$
hour-long integration times \citep{vanStraten:01,Jacoby:05},
corresponding to an improved dark matter mass detectability limit
of $\sim 10^{3} \, M_\oplus$ for these pulsars. There is an
exciting possibility \citep{TEMPO2} that pulsar timing may soon be
accurate to the $\mathcal{O}(1 \, \mathrm{ns})$ level, as the
software is no longer a barrier to such accuracy. Significant
improvements in measurement sensitivity would need to be made as
well, of course, but if that happens, it would allow for the
detectability of $\sim 10 \, M_\oplus$ objects, probing past the
expected cutoff in the mass function for WIMP dark matter.
Although it is estimated in \citet{TEMPO2} that a factor of $10^3$
improvement is needed in sensitivity to obtain timing accuracies
of $\sim 1 \, \mathrm{ns}$, the upcoming Square Kilometer Array
will have the sensitivity necessary to improve the residuals by up
to two orders of magnitude for many tens of pulsars.  It is not,
therefore, overly optimistic to consider the method of pulsar
timing as having the potential to detect dark matter.  We also
point out that the use of averaging techniques may allow for a
reduction in random errors in the pulsar residuals.  When the
Square Kilometer Array is complete, and routinely timing hundreds
of pulsars to sub-$\mu$s accuracy, it is possible that the
presence of dark matter clumps may limit the achievable accuracy
of pulsar timing, dependent upon the masses and number densities
of the dark matter clumps.

An accurate mass function can be derived from Press-Schechter
and/or Sheth-Tormen theory via equations (\ref{PS},\ref{ST}) and
the linear power spectrum and transfer function, which can be
obtained for any given model of dark matter. Realistically, the
mass function can be probed all the way down to this detectability
limit and perhaps further, if superior analysis techniques are
developed. We emphasize that probing dark matter substructure
within our own galaxy is the only handle on structure formation at
scales below $\sim 100 \, \mathrm{kpc}$ at present. Furthermore,
there may also be a correlation between dark matter microhalos
that are either sufficiently massive, dense, or nearby and a
possible WIMP annihilation signal \citep{EGRET:05}, which may
provide further information about the nature of dark matter.


Uncovering the nature of dark matter remains one of the most
important problems in cosmology today.  Precision pulsar timing
measurements can be used to probe dark matter substructure
transiting near the line-of-sight within our own galaxy.  Changes
in the light-travel-time of emitted pulses, although small (of
order $10^{-6} \, \mathrm{s}$) and over long $(\sim 10 \,
\mathrm{years})$ timescales, are nonetheless observable {\it in
practice} based on the current accuracy of pulsar timing.
Improvements in the understanding of pulsars themselves, as well
as the local environments of the pulsar and our solar system, will
allow us to probe smaller and smaller mass scales.  Additionally,
information gained about the transiting substructure can be used
to constrain both the particle properties of dark matter and the
density profiles of that structure.  Overall, pulsar timing
measurements may provide a new window into uncovering the
properties, method of creation, and evolution of the non-luminous
matter in our universe.


\section*{Acknowledgments}

We acknowledge the Les Houches Summer School for hospitality and
thank the Dark Matter Working Group and Edoardo Di Napoli in
particular for extremely helpful conversations during the
formative stages of this work. We also acknowledge Kathryn Zurek,
and Konstantin Borozdin for useful discussions. E.R.S. thanks the
University of Wisconsin for support during the completion of this
work.  M. P. H. is supported in part by funds provided by the U.S.
Department of Energy (D.O.E.) under cooperative research agreement
DE-FG02-05ER41360.


\end{document}